\documentclass[12pt]{article}
\pdfoutput=1
\usepackage{jheppub}
\usepackage{amsfonts}
\usepackage{amsmath,amssymb} 
\usepackage{graphicx} 
\usepackage[dvipsnames,x11names]{xcolor} 

\usepackage[nottoc,notlof,notlot]{tocbibind} 
\usepackage[titles]{tocloft} 
\usepackage{hyperref}
\usepackage{cleveref}
\usepackage{float}
\usepackage{subcaption}
\numberwithin{equation}{section} 
\usepackage{shuffle}
\usepackage{enumitem}

\title{\textbf{ETH-monotonicity and the black hole singularity}}

\author{Nilakash Sorokhaibam}\emailAdd{phy\_sns@tezu.ernet.in}

\affiliation{Department of Physics, Tezpur University, Tezpur, 784028, Assam, India.}

\abstract{We study the enveloping function of the fluctuation term in eigenstate thermalization hypothesis (ETH) statement for holographic conformal field theories. We use this function to identify and examine black hole microstates. We set down a set of desirable criteria for this function called ETH-monotonicity. It reinforces the Kelvin statement of the second law of thermodynamics over and above the universal entropic contribution. We show that higher-dimensional holographic conformal field theories possess ETH-monotonicity. Stronger contribution from ETH-monotonicity to the second law of thermodynamics is observed in smaller black hole microstates. It dominates other quantum fluctuations. It also measures the curvature at the horizon of the small black holes. In the smallest size limit, the black hole curvature singularity is constructed of microstates for which ETH-monotonicity starts competing with the entropic contribution. We expect that ETH-monotonicity will persist even in the ultimate quantum theory of gravity. Because it is a property of many-body quantum chaotic systems which becomes more prominent with decreasing system size, unlike other physical properties which are usually more predictable with increasing system size. Two-dimensional holographic conformal field theory does not possess all features of ETH-monotonicity which is in agreement with the absence of curvature singularity in the BTZ black hole.\\

\vspace{3.5cm}

\noindent{\it In a situation like this, discovering the right hypothesis is more than half the battle.}\\
\null \hfill --- James R. Munkres\\
\null \hfill in {\it Topology}}

\begin{document}
\maketitle

\section{Introduction}
\label{sec:intro}
It would not be an overstatement to state that the black hole singularity (and, in general, other spacetime singularities) has been the greatest mystery in physics. This singularity is found abundantly in physics. So, we expect that its physics underpinning should be robust and universal. Black holes are also known to be highly chaotic \cite{Sekino:2008he,Maldacena:2015waa}. In this work, we provide a promising lead towards resolving this mystery using a newly discovered property of quantum chaotic many-body systems. We show that this property arises from strong curvature at the black hole horizon.

We will concentrate on black holes in asymptotically Anti-de Sitter (AdS) spacetime. Black holes in asymptotically AdS spacetime corresponds to a thermal state on the holographic conformal field theory (CFT) at the AdS boundary by AdS/CFT correspondence \cite{Maldacena_1999,Witten:1998qj,Aharony_2000}. The temperature of the CFT is equal to the Hawking temperature of the black hole. We will connect certain property of the holographic CFTs with high curvature at the black hole horizon.

Quantum chaos and the related topic of thermalization in closed quantum systems have been subjects of great interest for the last few decades \cite{Polkovnikov:2010yn,DAlessio:2015qtq,Gogolin:2016hwy}. It is expected that closed chaotic quantum systems thermalize in the long time limit. This explains the validity of quantum statistical mechanics in nature. In this regard, eigenstate thermalization hypothesis (ETH) \cite{PhysRevA.43.2046,Srednicki:1994mfb} explains thermalization after unitary time evolution. So, ETH has been a subject of intense study in different set-ups, mostly in finite lattice systems using exact diagonalization \cite{DAlessio:2015qtq}.

ETH is concerned with few-body observables which can be easily measured or calculated. We say that a system is thermalized if these observables have expectation values equal to the thermal expectation values. Consider a non-fermionic hermitian operator $\mathcal{O}$ which corresponds to one such observable. ETH ansatz of the matrix elements $\mathcal{O}_{mn}$ in energy eigenstates is \cite{DAlessio:2015qtq}\footnote{We are assuming that the system obeys time-reversal invariance. Otherwise, $f(\bar{E},\omega)=f^*(\bar{E},-\omega)$ and $R_{nm}=R^*_{mn}$. It does not affect our calculations because all our quantities will be in terms of $|f(\bar{E},\omega)|$.}
\begin{gather}
\label{ETH}
\langle m|\mathcal{O}|n\rangle\equiv \mathcal{O}_{mn}=\mathcal{O}(\bar{E})\delta_{mn}+e^{-S(\bar{E})/2} f(\bar{E},\omega)R_{mn}\\
\bar{E}=(E_m+E_n)/2, \quad \omega=E_m-E_n, \quad f(\bar{E},\omega)=f(\bar{E},-\omega) \geq 0, \quad R_{nm}=R_{mn},\nonumber\
\end{gather}
where $|m\rangle$ and $|n\rangle$ are energy eigenstates with energies $E_m$ and $E_n$ respectively. $S(\bar{E})$ is the entropy at energy $\bar{E}$. $\mathcal{O}(\bar{E})$ and $f(\bar{E},\omega)$ are smooth functions of their arguments.  So, $\mathcal{O}(\bar{E})$ is equal to the expectation value of $\mathcal{O}$ in the microcanonical ensemble centered at energy $\bar{E}$. $R_{mn}$ are pseudo-random variables with zero mean and unit variance. At large $|\omega|$, the $f$-function falls at least as fast as $e^{-|\beta\omega|/4}$, where $\beta$ is the effective inverse temperature \cite{Srednicki_1999,DAlessio:2015qtq,Dymarsky:2018ccu,Murthy:2019fgs}. The second fluctuation term in (\ref{ETH}) exists even for the diagonal elements. The ETH statement for the diagonal elements without the fluctuations is equivalent to Berry's conjecture \cite{Berry_1977}. It implies that eigenstate expectation values are equal to the microcanonical averages in quantum chaotic systems \cite{DAlessio:2015qtq}.

The $\omega$-dependence of the enveloping function $f(\bar{E},\omega)$ controls the response function of the operator $\mathcal{O}$ \cite{Srednicki_1999}. So, the $\omega$-dependence has been studied for many systems \cite{DAlessio:2015qtq}. We concentrate on the $\bar{E}$-dependence in this work. The main technical advance made in this work is that \emph{we demonstrate the use of the $\bar{E}$-dependence of the $f$-function to identify and examine black hole microstates}. We use this technique to identify small black hole microstates. Then we identify a quantity of the holographic CFT which measures the strong curvature at the horizon of these small black holes. This quantity arises from inequivalence of ensembles so it increases with decreasing size of the black hole, thereby measuring the increasing curvature at the horizon of the black hole. We will write the $f$-function as $f(\bar{E})$ when $\omega$ is fixed and the $\omega$-dependence is not important. We calculate $f(\bar{E},\omega)$ from the spectral function. The spectral function is the imaginary part of the retarded Green function. The retarded Green function is calculated using the well-known Son-Starinets prescription \cite{Son:2002sd}.

Recently, the $\bar{E}$-dependence was studied in lattice models in \cite{ETHmono,ETHmono2D}. It is also closely related with the second law of thermodynamics. We found an interesting set of constraints on the $\bar{E}$-dependence of the $f$-function which we called ETH-monotonicity. In this work, we will explore ETH-monotonicity in holographic CFTs. We find a concrete connection between ETH-monotonicity and high-curvature at the black hole horizon. The present work is largely independent even though we got the inspiration from the above previous works. The lattice models considered in those works have various energy scales. The scaling of $f(\bar{E})$ with varying volume of the lattice model was studied. In the present work, we are studying conformal field theory so we look for scaling of $f(\bar{E})$ with varying $\bar{E}$.

ETH in CFTs has been studied in \cite{Lashkari:2016vgj,Basu:2017kzo,He:2017txy,Brehm_2018,Hikida_2018,Dymarsky:2019etq}. In particular, \cite{Brehm_2018,Hikida_2018} have shown the exponential suppression by the entropy for the off-diagonal elements in large-c CFTs. For the rest of the paper, we will work with the following assumptions:
\begin{enumerate}
\item Holographic CFTs satisfy ETH.
\item We will study $\bar{E}\to 0$ behaviour. This notation is for conciseness. We are more interested in the behaviour of the $f$-function as $\bar{E}$ decreases. We are still working with many-body excited states. In the AdS side, we are still working with the small black hole microstates in the Hilbert space of string theory \cite{Aharony_2000}. We are not considering stringy or quantum gravity corrections.
\end{enumerate}

It is well known that certain geodesics can explore the black hole singularity \cite{Fidkowski:2003nf,Festuccia:2005pi,Rodriguez-Gomez:2021pfh,Horowitz:2023ury,Ceplak:2024bja,Afkhami-Jeddi:2025wra,Dodelson:2025jff}. But the exact nature of the singularity has been as elusive as ever. The study of these ``bouncing" geodesics is related to the $\omega$-dependence of $f(\bar{E},\omega)$, while we are concerned with the $\bar{E}$-dependence. Most of these works also considered the simplified black brane geometry. Moreover, it has been found that a certain black brane geometry with curvature singularity does not have any bouncing geodesic \cite{Grozdanov:2026cut}. Other related works are \cite{Grinberg:2020fdj,David:2022nfn,Singhi:2024sdr,Jia:2025jbi}. We do not consider black branes because all black branes are related by a simple coordinate rescaling and all of them are already in the thermodynamic limit.

The outline of the rest of the paper is as follows. We review the necessary basics of AdS/CFT correspondence in section \ref{ads-cft}. ETH-monotonicity is defined and studied in section \ref{sec:ETHmono}. Section \ref{sumres} contains summary of our results and some discussions. Quantum fluctuations in small black holes are studied in section \ref{qf}. Section \ref{4d-CFT} contains our main results. We show that higher dimensional holographic CFTs possess ETH-monotonicity. The special case of two dimensional holographic CFT is studied in section \ref{2d-CFT}. Section \ref{cnd} consists of conclusions from this work. Microcanonical ensemble is studied in detail in Appendix \ref{app:mc}. Further numerical results including the case of 3-d holographic CFT are presented in Appendix \ref{app:num-res}.

For ease of reference, we list out the various energy scales and the various competing quantities in section \ref{cnd}.
\section{Black holes and AdS/CFT correspondence}
\label{ads-cft}
Black holes in global $(d+1)$-dimensional ($d+1$-D) AdS spacetime is described by the AdS-Schwarzchild metric
\begin{gather}
ds^2=-g(r)dt^2+g(r)^{-1}dr^2+r^2d\Omega_{d-1} \qquad g(r)=1-\frac{\mu}{r^{d-2}}+\frac{r^2}{R^2}
\end{gather}
$\mu$ is directly proportional to the black hole mass $M$ \cite{Horowitz:1999jd}.  The black hole mass and the horizon radius $r_h$ are given by
\begin{gather}
M=\frac{2\pi^{d/2}(d-1)\mu}{16\pi \Gamma(d/2) G_N}=\frac{\mu}{\tilde{G}_N} \qquad \mu=r_h^{d-2}\left(1+\frac{r^2_h}{R^2}\right),
\label{Mmudef}
\end{gather}
where $\Gamma$ is the Gamma function and $G_N$ is the Newton's constant. The horizon radius is the largest zero of $g(r)$. In this work, we will not consider black holes with charges or angular momentum. The Hawking temperature and the Bekenstein-Hawking entropy of the black hole are given in terms of $r_h$ by
\begin{equation}
T=\frac{dr^2_h+(d-2)R^2}{4\pi r_h R^2}, \qquad S=\frac{2\pi^{d/2}\,r^{d-1}_h}{4\,\Gamma(d/2)\,G_N}
\label{TSdef}
\end{equation}
In the rest of the paper, we will set the AdS radius $R=1$. For $D\geq 4$, $T$ has the global minimum at $r_{h,min}=\sqrt{(d-2)/d}$. $T$ increases as $r_h$ decreases for small black holes $r_h<r_{h,min}$. So, there are three possible phases at any given temperature - thermal AdS, small black hole, and large black hole. The small black holes have negative specific heat like black holes in asymptotically flat spacetime and they are always unstable in the canonical ensemble. In other words, canonical ensemble is ill-defined for small black holes in $D\geq 4$ spacetime due to the superlinear entropy growth with increasing energy,
\begin{gather}
S(\bar{E})\sim (\bar{E}l_P)^{(d-1)/(d-2)}, \qquad d=D-1.
\end{gather}
$l_P=G_N^{1/(d-1)}$ is the Planck length scale. The stable phases are the AdS thermal gas and the large black hole phase which are separated by a first order Hawking-Page transition \cite{Hawking:1982dh}. The critical temperature is given by $r_h=1$ which is slightly away from $r_{h,min}$. But the small black holes are stable in the microcanonical ensemble. As such closed systems, they are also stable from perturbations \cite{Regge:1957td}. Dynamically, their lifetimes are very long even in the canonical ensemble \cite{Horowitz:1999uv}. Small charged black holes are unstable from perturbations \cite{Cardoso:2004hs,Cardoso:2006wa,Basu:2010uz}, the end-points are still small hairy black holes. For $D=3$, the Hawking-Page transition is still between the thermal AdS phase and the black hole phase at the critical point $r_h=1$. The BTZ black hole \cite{Banados:1992wn} in $D=3$ always have a positive specific heat. We are not interested in black strings and black branes so we are not concerned with Gregory-Laflamme instability \cite{Gregory:1993vy}.

We would still be working with small black holes. In terms of thermodynamics, we will work in microcanonical ensemble. Small black holes are of particular interest for the present work because small black holes have high curvature at the horizon. Note that in the other limit $r_h\to\infty$, canonical ensembles are well defined in any dimensions.

By AdS/CFT correspondence, black holes in the $(d+1)$-D spacetime corresponds to a thermal state on the $d$-dimensional ($d$-d) holographic conformal field theory (CFT) at the AdS boundary. The temperature of the CFT is equal to the Hawking temperature of the black hole. The total energy of the CFT system $\bar{E}_{\beta}$ is equal to the mass of the black hole $M=\mu/\tilde{G}_N$ \cite{Balasubramanian:1999re}. From the definition of $\bar{E}=(E_n+E_m)/2$ and $\omega=E_m-E_n$, $|\omega|\leq 2\bar{E}$. But we do not have access to the discrete energy spectrum of quantized gravity. $\bar{E}$ (energy of \emph{heavy} operators) is enhanced by a $1/G_N$ factor as compared to $\omega$ (energy of \emph{light} operators) when measured in the unit of $1/R$. Hence, we can effectively take $\omega\in(-\infty,\infty)$. The use of these two different energy scales were implicit in other works, for example, see \cite{Horowitz:1999jd,Jokela:2015sza}.
 
In this work, we will consider perturbation with a light operator on the CFT. We are working with linear response theory so the strength of the source is chosen to be small. It has been shown that strong global perturbation with a light operator can form black holes \cite{Bhattacharyya:2009uu}. One can use geodesic approximation \cite{Balasubramanian:1999zv} in case of perturbation with heavy operators. The technical challenge is that there are no real geodesic which connects two timelike boundary points \cite{Balasubramanian:2012tu}.

We use the spectral function $A(\beta,\omega)$ of a scalar operator in the holographic CFT to calculate $f(\bar{E})$ of the scalar operator. In a thermal state with inverse temperature $\beta$ and energy expectation value $\bar{E}_{\beta}$,
\begin{gather}
\label{ETHspec}
A(\beta,\omega)=2\sinh(\beta\omega/2)f(\bar{E}_{\beta},\omega)^2, \qquad A(\beta,\omega)=-2\,\text{Im}\;\tilde{G}_R(\beta,\omega)\\
G_R(t,t')=-i\theta(t-t')\langle[\mathcal{O}(t),\mathcal{O}(t')]\rangle_{\beta}
\end{gather}
The retarded Green function $\tilde{G}_R(\beta,\omega)$ is calculated using the Son-Starinets prescription \cite{Son:2002sd}. We consider a minimally-coupled massive scalar field $\phi$ of mass squared $m^2$ in the $d+1$-D bulk spacetime. The equation of motion in terms of the Fourier modes of $t$ and the spherical harmonics of the $(d-1)$-sphere is
\begin{equation}
\left(\frac{1}{r^{d-1}}\partial_r\left(r^{d-1}g(r)\partial_r\right)+\frac{\omega^2}{g(r)}-\frac{l(l+d-2)}{r^2}-m^2\right)\phi_{\omega l}=0,
\label{KGeqn}
\end{equation}
where $\omega$ is the frequency and $l$ is the angular momentum mode number of the spherical harmonics. By AdS/CFT correspondence, this scalar field is dual to a scalar quantum operator $\mathcal{O}_{\Delta+}$ of dimension $\Delta_+$ in the holographic CFT. There could be another operator $\mathcal{O}_{\Delta-}$ of dimension $\Delta_-$ for a certain range of $m^2$. $\Delta_{\pm}$ is a function of $m$. $m$ has a lower bound fixed by the Breitenlohner-Freedman (BF) bound \cite{Breitenlohner:1982jf}
\begin{equation}
\Delta_{\pm}=\frac{d}{2}\pm\sqrt{\frac{d^2}{4}+m^2}, \qquad m^2>-\frac{d^2}{4} \quad \text{(BF bound)}
\end{equation}
The scalar operator $\mathcal{O}_{\Delta-}$ is well defined \cite{Klebanov:1999tb} for the mass range
\begin{equation}
-\frac{d^2}{4}<m^2<-\frac{d^2}{4}+1 \quad \Rightarrow \quad \Delta_{-}>\frac{d-2}{2} \quad \text{(Unitary bound)}
\end{equation}
We impose ingoing boundary condition at the horizon. The coefficients of the independent solutions at the boundary gives us the retarded Green function \cite{Son:2002sd}.
\begin{gather}
\phi_{\omega l}(r \to r_h)=(r-r_h)^{-i\omega/4\pi T}+...\\
\phi_{\omega l}(r \to \infty)= \mathcal{A}(\omega,l)r^{-\Delta_-}\left(1+O(r^2)\right)+\mathcal{B}(\omega,l)r^{-\Delta_+}\left(1+O(r^2)\right)\\
\qquad \tilde{G}_{R,\Delta+}(\beta,\omega)=-\frac{\mathcal{B}(\omega,l)}{\mathcal{A}(\omega,l)}
\label{SonStarinets}
\end{gather}
For brevity, we have suppressed the $l$-dependence in $\tilde{G}_{R,\Delta+}(\beta,\omega)$, and instead, emphasized $\beta$-dependence. If $\mathcal{O}_{\Delta-}$ is well defined, then its retarded Green function is given by
\begin{equation}
\tilde{G}_{R,\Delta-}(\beta,\omega)=\frac{\mathcal{A}(\omega,l)}{\mathcal{B}(\omega,l)}
\label{AbyBformula}
\end{equation}
The $f$-function can be calculated using (\ref{ETHspec}), where we have suppressed the $l$-dependence. We are looking for $\mu$, $\beta$, $T$ or $r_h$ dependence. While in other works, $\omega$-dependence is the priority. So it is important that (\ref{SonStarinets}) is nicely normalized $\text{lim}_{t\to 0+} G_R(t)=-i/t^{2\Delta}$, while using the $r$ coordinate \cite{Dodelson:2022yvn}. It should be noted that a different coordinate system was used in \cite{Son:2002sd}. For only the spectral function, we can also perform the calculation near the horizon due to flux conservation. In the rest of the paper, we will consider an operator $\mathcal{O}_{\Delta}$ of conformal weight $\Delta$ which can be either $\Delta_+$ or $\Delta_-$.

We will be varying $\mu$ so we are really working with $f(\bar{E})=f(\mu/\tilde{G}_N)=\hat{f}(\mu)$. With a slight abuse of notation, we will still use $f(\mu)$ instead of $\hat{f}(\mu)$. The quantity that appears in physical processes is $f'(\bar{E})/f(\bar{E})$. So we will use the relation
\begin{equation}
\frac{f'(\bar{E})}{f(\bar{E})}=\tilde{G}_N\,\frac{f'(\mu)}{f(\mu)}
\label{GNsup}
\end{equation}

\section{ETH-monotonicity}
\label{sec:ETHmono}
The Kelvin statement of the second law of thermodynamics implies that an isolated system always gains energy when perturbed momentarily \cite{Kardarbook2007,huang2008statistical}. It is related to the well-known Planck statement of the second law via the relation
\begin{equation}
\Delta E=T\Delta S
\end{equation}
$\Delta E$, $T$ and $\Delta S$ are change in energy, temperature and change in entropy. $\Delta S$ is positive when $\Delta E$ has the same sign as $T$.

$f(\bar{E})$ is closely related with the second law of thermodynamics. In \cite{ETHmono,ETHmono2D}, we found some interesting constraints on $\bar{E}$-dependence of the $f$-function in (\ref{ETH}). These constraints are collectively called ETH-monotonicity. These constraints arise from a non-trivial relation between Berry's conjecture and the second law of thermodynamics in the form of the Kelvin statement. It implies that, when the system is perturbed starting from an energy eigenstate or a microcanonical ensemble, more energy is absorbed than an equivalent canonical ensemble. The systems considered were lattice models with various energy scales. So, we will provide motivation and redefine ETH-monotonicity for holographic CFTs below. The main implication is that \emph{an energy eigenstate absorbs more energy than an equivalent canonical ensemble, when the system is perturbed}. The low energy eigenstates are of particular interest. The high energy eigenstates gain energy same as their equivalent canonical ensembles due to equivalence of ensembles.

Now we will define ETH-monotonicity for holographic CFTs. It is a set of desirable criteria for the function $f(\bar{E})$. We will also provide the motivations for these criteria. ETH-monotonicity for holographic CFTs are the following criteria:
\begin{enumerate}[label=R{{\arabic*}}.]
\item $f(\bar{E})$ is locally a constant function of $\bar{E}$ as $\bar{E}\to\infty$.
\item $f(\bar{E})$ is a monotonically increasing function of $\bar{E}$.
\item $f(\bar{E})$ stiffens in $\bar{E}\to 0$ and flattens in $\bar{E}\to\infty$ such that
\begin{gather}
\label{ETHmono0}
0\;<\;\lim_{\mu\to 0}\mu^{\eta}\,\frac{f'(\mu)}{f(\mu)}=\lim_{\bar{E}\to 0}\tilde{G}^{\eta-1}_N\bar{E}^{\eta}\,\frac{f'(\bar{E})}{f(\bar{E})}=\kappa\;<\;\infty, \quad \eta=1\\
0\;<\;\lim_{\bar{E}\to \infty}\bar{E}^{\nu}\,\frac{f'(\bar{E})}{f(\bar{E})}\;<\;\infty, \qquad \nu>0
\label{ETHmonoinf}
\end{gather}
\end{enumerate}
(\ref{ETHmono0}) is the most severe constraint, $\eta=1$ is necessary. We emphasize that while taking $\bar{E}\to 0$ we are not shrinking the horizon radius to the string length scale $l_s$ or the Planck length scale $l_P$. In other word, the ground state $\bar{E}=0$ corresponds to the smallest black hole in the Hilbert space of string theory. We will briefly comment on the quantum gravity limit in the next section \ref{sumres}. $\bar{E}=0$ is the ground state because the Casimir energy does not participate in the dynamical set-up that we are considering. All CFTs do not have any intrinsic system size scale. So, R3 is formulated in terms of $\bar{E}$. R3 was originally formulated in terms of the degrees of freedom $L^d$, where $L$ is the linear system size and $d$ is the spatial dimension of the system.

For the Kelvin statement of the second law, the perturbation is by a source term $\lambda(t)\mathcal{O}$. Starting from a canonical ensemble with energy $\bar{E}_{\beta}$, the gain in energy by the system at leading order in perturbative expansion is \cite{ETHmono}
\begin{gather}
\text{Entropic contribution:} \qquad \Delta E_{\beta}=\int_{-\infty}^{\infty}d\omega\, \omega |\tilde{\lambda}(\omega)|^2\,e^{\beta\omega/2} f(\bar{E}_{\beta},\omega)^2 \quad \sim \quad O(1)
\label{thermalDE}
\end{gather}
$\tilde{\lambda}(\omega)$ is the Fourier transform of $\lambda(t)$. For a hermitian operator, $\lambda(t)$ is real and $\tilde{\lambda}(\omega)=\tilde{\lambda}(-\omega)^*$. The above quantity is identically positive (negative or zero) for positive (negative or infinite) temperature. So, the second law of thermodynamics is guaranteed in canonical ensembles (in general, for any \emph{passive} density matrix) \cite{Polkovnikov_2008}. The integral is finite because the $f$-function falls at least as fast as $e^{-|\beta\omega|/4}$, otherwise the theory is not UV complete. $\Delta E_{\beta}$ is of order $O(1)$ because we are perturbing the system with a light operator. Note that we have solved the backreacted geometry at leading order in perturbation theory. The gain in energy is the change in the mass of the black hole. We are performing the calculation from the CFT side. It is possible because of our assumption that holographic CFTs satisfy ETH.

We will call $\Delta E_{\beta}$ the entropic contribution to the gain in energy or to the second law. We do not want to call it the change in energy starting from a canonical ensemble because canonical ensemble is ill-defined for small black holes. The $e^{\beta\omega/2}$ factor comes from the competition of entropies, as we will see soon. Moreover, this contribution is solely responsible for the second law in the thermodynamic limit.
 
Starting from an energy eigenstate $E_n$, the gain in energy is given by
\begin{eqnarray}
\label{microDEraw}
\Delta E_{n}&=&\sum_m (E_m-E_n)|\tilde{\lambda}(E_m-E_n)|^2|\mathcal{O}_{mn}|^2+O(\tilde{\lambda}^4)\\
\label{microDEexact}
&=&\int_{-\infty}^{\infty}d\omega\, \omega |\tilde{\lambda}(\omega)|^2\,e^{S(E_n+\omega)-S(E_n+\omega/2)} f(E_n+\omega/2,\omega)^2\\
&=&\int_{-\infty}^{\infty}d\omega\, \omega |\tilde{\lambda}(\omega)|^2\,e^{\beta\omega/2} f(E_{n}+\omega/2,\omega)^2
\label{microDE}
\end{eqnarray}
We have used Taylor expansions of $S(E_n+\omega)$ and $S(E_n+\omega/2)$ about the energy of the initial state $E_n$, $\omega=E_m-E_n$, and $\beta=S'(\bar{E})|_{\bar{E}=E_n}$ is the effective inverse temperature at $E_n$. The difference with the expression of $\Delta E_{\beta}$ is in the $\bar{E}$-dependence. For a finite system (not thermodynamic limit), we will find that $f(\bar{E})$ is a monotonically increasing function of $\bar{E}$. It is also the criteria R2 that we set down in ETH-monotonicity. So the system gains more energy if the initial state is an eigenstate, over and above the entropic contribution. This extra gain in energy is due to ETH-monotonicity. We have neglected the contributions of the higher derivative terms $S''(\bar{E})$, $S'''(\bar{E})$, etc. They are suppressed by increasing positive powers of $G_N$ factor. But $S''(\bar{E})$ can compete with ETH-monotonicity. This term will be simply called quantum fluctuation. We will study it in details in section \ref{qf}. These fluctuations, from both $f(E_{n}+\omega/2,\omega)$ and $S''(\bar{E})$, have been observed earlier (see around eqn (211) in \cite{DAlessio:2015qtq}). They were called ``fluctuation within each eigenstate". But the functional form of $f(\bar{E})$ has not been studied anywhere before \cite{ETHmono,ETHmono2D}. The ``fluctuation from energy fluctuations" is the microcanonical fluctuations $\langle \mathcal{H}_0^2\rangle-\langle \mathcal{H}_0\rangle^2$ measurable in experiments. $\mathcal{H}_0$ is the Hamiltonian of the system. This fluctuation is zero for an energy eigenstate, but non-zero for a microcanonical ensemble. So either we work with an energy eigenstate or simply concentrate on expectation value of the energy when working with a microcanonical ensemble.

(\ref{microDE}) also holds true in a microcanonical ensemble with energy expectation value $\bar{E}_{mc}$, see Appendix \ref{app:mc} for details. The microcanonical energy window $\delta$ is of the order of $O(1/\sqrt{G_N})$. But $\delta$ is usually taken to be parametrically much smaller than $\sqrt{G_N}\bar{E}_{mc}\sim 1/\sqrt{G_N}$ \cite{Srednicki_1999}. The correction from the microcanonical energy window $\delta$ arises from two different terms - one from the entropy exponentials and the other from $f(\bar{E})$. But both corrections are suppressed by $e^{-1/G_N}$. So $\delta$ does not affect the extra gain in energy from ETH-monotonicity in any dimension. This is expected because ETH-monotonicity is one of the manifestations of inequivalence of ensembles and we are working with microcanonical ensemble. The energy fluctuation $\langle \mathcal{H}_0^2\rangle-\langle \mathcal{H}_0\rangle^2 \sim \delta \sim O(1/\sqrt{G_N})$ is significant for experimental measurements. As we will see, we are looking for $O(G^1_N)$ effects which are very small compared to this energy fluctuation in measurements. So, one has to perform a large number of very high precision energy measurements to discern the $O(G^1_N)$ effects (no free lunch). But we will concentrate on the expectation value and ignore the measurement fluctuations. In other words, our priority is to keep track of the minute $O(G^1_N)$ effects theoretically. The microcanonical energy fluctuation is simply an experimental challenge.

In holographic CFTs, $\bar{E}$ is order $1/\tilde{G}_N$ times larger than $\omega$. So, we can Taylor expand $f(E_{n}+\omega/2,\omega)$ about $(E_{n},\omega)$ in (\ref{microDE}). Using the Taylor expansion, we can extract the extra gain in energy from ETH-monotonicity when the system is initially in a microcanonical ensemble. Let us denote that quantity by $\Delta E_{mc,f}$ and it is given by
\begin{gather}
\label{taylorexp}
\Delta E_{mc} = \int_{-\infty}^{\infty} d\omega \, \omega|\tilde{\lambda}(\omega)|^2 e^{\beta\omega/2}f(\bar{E}_{mc},\omega)^2\left[1+\frac{\omega}{f(\bar{E}_{mc},\omega)}\left.\frac{\partial f(\bar{E},\omega)}{\partial\bar{E}}\right|_{\bar{E}=\bar{E}_{mc}}\right]\\
\Delta E_{mc,f} = \int_{-\infty}^{\infty} d\omega \, \omega^2|\tilde{\lambda}(\omega)|^2 e^{\beta\omega/2}f(\bar{E}_{mc},\omega)\left.\frac{\partial f(\bar{E},\omega)}{\partial\bar{E}}\right|_{\bar{E}=\bar{E}_{mc}} \quad \sim \quad O(G^1_N)
\label{deltaE_ETHmono}
\end{gather}
The extra gain in energy is order $O(G^1_N)$ because it is controlled by $f'(\bar{E})/f(\bar{E})$, compared to $\Delta E_{\beta}$. So $f'(\bar{E})/f(\bar{E})$ is the main object of our interest.

{\bf Pure states:} In place of the microcanonical ensemble, one can also consider pure states $|\Psi\rangle=\sum_n c_n|n\rangle$ with $|n\rangle$ within the energy window $\delta$ of the order $O(1/\sqrt{G_N})$. These states have been called typical states in \cite{Srednicki_1999}. For the change in energy, the calculation is same as in the microcanonical ensemble because the cross terms $\mathcal{O}_{n_1m}\mathcal{O}_{mn_2}$ with $n_1\neq n_2$ in (\ref{microDEraw}) drops out due to the pseudo random factor $R_{n_1m}R_{mn_2}$ averaging out to zero.

The motivations for ETH-monotonicity are as follows. R1 follows from equivalence of ensembles in the thermodynamic limit $\bar{E}\to\infty$. (\ref{ETHmonoinf}) of R3 is the mathematical expression. It implies $\Delta E_{mc}\to\Delta E_{\beta}$ as a power law in $\bar{E}_{mc}$ when taking the thermodynamic limit. $f(\bar{E})$ quantifies fluctuations (not relative fluctuations) at the energy $\bar{E}$. So, it is expected to be an increasing function of $\bar{E}$. This gives R2. Many observables in chaotic lattice models have been found to obey R2 in \cite{ETHmono,ETHmono2D}.

We have special interest in the limit $\bar{E} \to 0$ because it corresponds to small black holes with high curvature at the horizon, and ultimately the black hole singularity. Naively, we expect that the extra gain in energy would persist in the limit $\bar{E}\to 0$ if $\eta>0$ in (\ref{ETHmono0}). $f'/f$ will diverge as $\kappa\bar{E}^{-\eta}$. It ensures that $\Delta E_{mc,f}$ increases compared to $\Delta E_{\beta}$ (or $\Delta E_{mc,f}/\Delta E_{\beta}$ increases). For example, we will find that $2-d$ holographic CFT has $\eta=3/2, \kappa=\pi\omega/4$. But with this value of $\eta$, $f(\bar{E})$ is
\begin{equation}
\mu^{3/2}\,\frac{f'(\mu)}{f(\mu)}=\tilde{G}^{1/2}_N\bar{E}^{3/2}\,\frac{f'(\bar{E})}{f(\bar{E})}=\kappa \quad \Rightarrow\quad f(\bar{E}) = f_0 e^{-2\kappa/\sqrt{\tilde{G}_N\bar{E}}}, \quad \kappa=\frac{\pi\omega}{4}
\end{equation}
where $f_0>0$ is the integration constant. The problem with this function is that $\Delta E_{mc}, \Delta E_{\beta}\sim e^{-1/\bar{E}}$ vanishes exponentially fast. There could be power-law dependent factor of $\bar{E}$ but it is not important for $f'(\bar{E})/f(\bar{E})$ in the presence of the above exponential. On the other hand, if $\eta=1$, then $f(\bar{E})$ is
\begin{equation}
f(\bar{E}) = f_0 \bar{E}^{\kappa}.
\label{fE0}
\end{equation}
So, $f(\bar{E})$ vanishes as a power law in $\bar{E}$. Another useful consequence is that $\kappa$ is generally expected to be independent of $\omega$, on general consideration. In the present physics problem, we do not expect a function like $\bar{E}^{\beta\omega}$ or $\bar{E}^{\log (\beta\omega)}$ where $\bar{E}$ is the energy of the system and $\omega$ is the frequency of the perturbation or linear response. This remark on $\kappa$ will have an interesting consequence later on.

If $\eta$ is less than $1$, say $\eta=1/2$, then $f(\bar{E}) = f_0 e^{2\kappa \sqrt{\tilde{G}_N\bar{E}}}$ which is not possible because $\bar{E}=0$ is the ground state. $f(\bar{E})$ signifies the strength of quantum fluctuations mimicking thermal fluctuations \cite{Srednicki_1999} and it is expected to vanish for the ground state. So in conclusion, $\eta=1$ is a very strict constraint in (\ref{ETHmono0}).

Note that $\eta>0$, in particular $\eta=1$, implies that ETH-monotonicity can start competing with the entropic contribution in the limit $\bar{E}\to 0$. We can examine this competition by calculating the relative extra gain in energy at a particular frequency $\omega$ defined by
\begin{equation}
\overline{\Delta E}_{mc}=\frac{\Delta E_{mc,f}}{\Delta E_{\beta}}=\frac{\omega \,\coth(\beta\omega/2)}{f(\bar{E},\omega)}\,\left.\frac{\partial f(\bar{E},\omega)}{\partial\bar{E}}\right|_{\bar{E}=E_{mc}} \quad \sim \quad O(G^1_N)
\label{Ediffratio}
\end{equation}
Note that $-\omega$ also contributes. Because when we perturb the system with a source at frequency $\omega$, the transition amplitudes to the eigenstate at $E_{mc}-\omega$ and the eigenstate at $E_{mc}+\omega$ are both non-zero. The competition gives rise to the second law of thermodynamics. The extra gain in energy is very small, suppressed by a $\tilde{G}_N$ factor from (\ref{GNsup}). So, black hole microstates are indeed highly thermal in nature. But we will see that this minute gain in energy can be analysed and $\overline{\Delta E}_{mc}$ is larger for smaller black holes which have stronger curvature at the horizon.

The frequency $\omega$ is fixed in the above argument. For simplicity, we will even choose a perturbation which is predominantly of a certain value of $\omega$. Consider the case of shooting a laser pulse on an electronic system. We are fixing the frequency of the laser pulse irrespective of the system size or the system temperature.

\section{Summary of results}
\label{sumres}
We use the $\bar{E}$-dependence of $f(\bar{E},\omega)$ in the ETH statement (\ref{ETH}) to identify and examine black hole microstates. In particular, we are interested in the microstates of small black holes which have high curvature at the horizon. We set down a set of criteria on the $\bar{E}$-dependence which we called ETH-monotonicity. Our technical results are:
\begin{enumerate}
\item Higher-d holographic CFTs possess ETH-monotonicity. Our numerical results suggest that $\eta=\nu=1$. Moreover, we find that $\kappa=\kappa_0+d_{\kappa}l$ with $\kappa_0=0.5$, where $d_{\kappa}$ is a fixed positive number and $l$ is the angular momentum mode. From the numerical results of the $3$-d system, we found that $\kappa=(1+2l)/2$. In the $4$-d system, $\kappa=(1+l)/2$. We can also extract the analytic expression of $f(\bar{E})$ for $l=0$ from the small $\mu$ expansion in \cite{Dodelson:2022yvn}. It is given by
\begin{equation}
\lim_{\bar{E}\to 0} f(\bar{E},\omega)=\sqrt{\frac{\tilde{G}_N \,\sin (\pi  \Delta )\, \Gamma (2-\Delta )\, \Gamma \left(\frac{\Delta -\omega }{2}\right) \,\Gamma \left(\frac{\Delta +\omega }{2}\right)}{\left[\cos (\pi  \Delta )-\cos (\pi  \omega )\right]\Gamma (\Delta -2)\Gamma \left(\frac{2-\Delta -\omega}{2}\right) \Gamma \left(\frac{2-\Delta +\omega}{2}\right)}} \; \bar{E}^{1/2}
\label{fl0exp}
\end{equation}

\item ETH-monotonicity is still prominent in a microcanonical ensemble defined in an energy window $\delta$ of the order $O(1/\sqrt{G}_N)$. In Higher-d holographic CFTs, the relative extra gain in energy $\overline{\Delta E}_{mc}$ measures the curvature at the black hole horizon of small black holes,
\begin{equation}
\lim_{\bar{E}\to 0}\overline{\Delta E}_{mc} = \frac{\kappa(d-2)\,\tilde{G}_N}{2\pi r_h^{d-1}} =\frac{\kappa(D-3)\,\tilde{G}_N}{2\pi r_h^{D-2}}, \quad \text{with} \quad D=d+1\geq 4.
\label{Ediffratioval}
\end{equation}
ETH-monotonicity also dominates the quantum fluctuation in the small black hole microstates.
\item $2$-d holographic CFT does not possess all the features of ETH-monotonicity. In particular, $\eta=3/2, \kappa=\pi\omega/4$ for $\Delta$ equal to integers and half integers. So, $f(\bar{E}) = f_0 e^{-2\kappa/\sqrt{\tilde{G}_N\bar{E}}}$ and the energy gains $\Delta E_{\beta}$ and $\Delta E_{mc}$ vanish exponentially as $\bar{E}\to 0$. Moreover, the quantum fluctuation negates and numerically dominates ETH-monotonicity in $2$-d holographic CFT.
\item For small black holes in higher dimensions, ETH-monotonicity is locally violated due to the appearance of quasiparticles. These quasiparticle peaks do not appear in case of $2$-d holographic CFT.
\end{enumerate}

The above results lead to our main conclusion that
\begin{center}
\emph{smaller black hole microstates in higher dimensions have stronger contribution from ETH-monotonicity to the second law of thermodynamics. In the smallest size limit, the black hole curvature singularity is constructed of microstates for which ETH-monotonicity starts competing with the entropic contribution. In summary,}\\
\vspace{0.2cm}
\bf{strong ETH-monotonicity = strong curvature at the horizon.}
\end{center}

The competition of ETH-monotonicity and the entropic factor is quantified by the relative extra gain in energy $\overline{\Delta E}_{mc}$. For small black holes,
\begin{equation}
E_{mc}=\frac{r_h^{d-2}}{\tilde{G}_N}, \qquad \beta=\frac{4\pi r_h}{d-2},\qquad \frac{f'(\bar{E})}{f(\bar{E})}=\tilde{G}_N\,\frac{f'(\mu)}{f(\mu)}=\frac{\kappa}{\bar{E}_{mc}}.
\label{smallBHval}
\end{equation}
Substituting these values in (\ref{Ediffratio}), we get (\ref{Ediffratioval}). The negative specific heat of small black holes implies that $\omega\coth(\beta\omega/2)\sim 2/\beta$ in (\ref{Ediffratio}) as $\beta\to 0$. $\kappa$ is also independent of $\omega$ in these higher-d CFTs. So the expression of $\overline{\Delta E}_{mc}$ is fixed by dimensional analysis after the $\omega$-dependence dropped out. As we have remarked above, we expect that $\kappa$ is independent of $\omega$ if $\eta=1$, or in general, if ETH-monotonicity is satisfied. We expect that this result would generalize to other black hole geometries with negative specific heat and satisfying ETH-monotonicity.

$\overline{\Delta E}_{mc}$ is proportional to the square root of the Kretschmann scalar $\sqrt{K}=\sqrt{R^{\mu\nu\rho\sigma}R_{\mu\nu\rho\sigma}}$ in $4$-D black holes. In higher-D black holes, $\overline{\Delta E}_{mc}$ measures higher order curvature invariants. The Kretschmann scalar at $r=r_h\to 0$ in different dimensions are
\begin{equation}
K=24+\frac{12\mu^2}{r_h^6}\to\frac{12}{r_h^4} \quad \text{in (4-D), \; and}\qquad K=40+\frac{72\mu^2}{r_h^8}\to\frac{72}{r_h^4} \quad \text{in (5-D).}
\end{equation}

It is worthwhile to note that if we naively take $r_h$ to the Planck length scale $l_P=G_N^{1/(d-1)}$, $\overline{\Delta E}_{mc}$ is of the order of one. So naively, the entropic contribution $\Delta E_{\beta}$ and the contribution from ETH-monotonicity become equal. One has to deal with quantum gravity at the Planck length scale. But we expect that ETH-monotonicity will persist even at this length scale, irrespective of the nature of quantum gravity. It is a property of many-body quantum chaotic system which becomes more prominent with decreasing system size, unlike other physical properties which are usually more predictable with increasing system size. It is generally agreed that the ultimate theory of quantum gravity will be a highly interacting chaotic theory. We have prominently observed ETH-monotonicity even in systems with 12 to 25 qubits \cite{ETHmono,ETHmono2D}.

In hindsight, our result is not surprising. Because canonical ensemble is ill-defined for small AdS black holes, as for black holes in flat spacetime. The entropy $S(\bar{E})$ grows super-linearly with increasing energy $\bar{E}$. So, one has to work with microcanonical ensemble. Then, the overarching idea is to exploit the inequivalence of ensembles to identify geometry with strong curvature. This is possible because of the surprising result that the extra gain in energy in microcanonical ensemble is always positive. We emphasize that our results are neither conjectures nor proposals. They follow from precise and concrete calculations. We have analysed all competing terms at order $O(G^1_N)$.

A large ``white hole"\footnote{\label{whitehole} We do not mean the white hole in the Penrose diagram of an eternal black hole. We mean the microstates in the upper half of the finite energy spectrum of quantum gravity. The opposite of small black holes are not small ``white holes". They are the highest energy microstates with small entropy, so large ``white holes".} is again a microstate with ETH-monotonicity dominance but at negative inverse temperature. A small perturbation will lead to the lost of an  excess amount of energy of the system, more than what one expects from only entropic consideration. The entropy will decrease with increasing energy for such ``white holes". So, $f(\bar{E})$ will be a decreasing function of $\bar{E}$ in this region of the energy spectrum.

The spectral functions and quasinormal modes in small AdS black holes have been studied in \cite{Horowitz:1999jd,Zhu:2001vi,Konoplya:2002zu,Berti:2009wx,Jokela:2015sza}. In particular, \cite{Jokela:2015sza} showed that the quasinormal modes and the spectral functions gradually tend towards that of the discrete modes in thermal AdS spacetime.

\section{Quantum fluctuations}
\label{qf}
The Taylor expansions of $S(E_n+\omega)$ and $S(E_n+\omega/2)$ in (\ref{microDEexact}) have higher derivative terms $S''(\bar{E})$, $S'''(\bar{E})$, etc. $S''(\bar{E})$ gives a correction which is of the order $O(G^1_N)$. $S'''(\bar{E})$ and higher derivative terms are suppressed by higher positive powers of $G_N$. So, we will concentrate on the contribution from $S''(\bar{E})=-\beta^2/C$, where $C$ is the total heat capacity \cite{Murthy:2019fgs}. It is due to quantum fluctuations which mimic thermal fluctuations. So far, we have been ignoring this term. We expand the exponential in powers of $G_N$. The leading contribution to the change in energy due to the quantum fluctuations is given by
\begin{equation}
\Delta E_{mc,q}=\frac{3}{4}\,\omega^3 |\tilde{\lambda}(\omega)|^2\,\sinh(\beta\omega/2) f(\bar{E},\omega)^2 S''(\bar{E})|_{\bar{E}=E_{mc}} \quad \sim \quad O(G^1_N).
\label{microDEqf}
\end{equation}
It can compete with the contribution from ETH-monotonicity $\Delta E_{n,f}$. We compare the contribution from the quantum fluctuations with the contribution from ETH-monotonicity. We calculate the quantity $\Delta E_{mc,q}/\Delta E_{\beta}$ in the limit $\bar{E}\to 0$ or $r_h\to 0$.
\begin{equation}
\frac{\Delta E_{mc,q}}{\Delta E_{\beta}}=\frac{3}{8}\,\omega^2 S''(\bar{E})|_{\bar{E}=E_{mc}}=\frac{3}{2}\left(\frac{d-1}{d-2}\right)^2\frac{\omega^2\tilde{G}_N}{\mu^{\frac{d-3}{d-2}}} = \frac{3}{2}\left(\frac{d-1}{d-2}\right)^2\frac{\omega^2\tilde{G}_N}{r^{d-3}_h}
\label{Ediffratioq}
\end{equation}
We have used $S(\bar{E})=4(d-1)\mu^{(d-1)/(d-2)}/\tilde{G}_N$ and $\bar{E}=\mu/\tilde{G}_N$. We can see that $\overline{\Delta E}_{mc}$ in (\ref{Ediffratioval}) grows faster that the above quantity as $r_h\to 0$. So, the contribution from ETH-monotonicity dominates the contribution from the quantum fluctuations for small black hole microstates in higher dimensions. What is interesting about $\Delta E_{mc,q}$ is that it is still a positive quantity so it further reinforces the Kelvin statement of the second law. This is tied to the fact that small black holes in higher dimensions have negative specific heat.

The behaviour of $\Delta E_{mc,q}/\Delta E_{\beta}$ for $2$-d holographic CFT is drastically different and will be considered in section \ref{2d-CFT}. We will find that the quantum fluctuations negate and numerically dominate ETH-monotonicity. Both their contributions are already exponentially small due to the $f(\bar{E})$ given by $\eta=3/2$.

\section{ETH-monotonicity in higher-d holographic CFT}
\label{4d-CFT}
In this section, we show that the higher-d holographic CFTs possess ETH-monotonicity. We calculate the spectral function numerically by solving the Klein-Gordon (KG) equation (\ref{KGeqn}) in AdS$_5$ black hole background. The retarded Green function has been calculated in \cite{Dodelson:2022yvn} as instanton expansion. It is still most straightforward to solve the KG equation numerically using NSolve in Mathematica. The Mathematica codes are available as an ancillary file at https://arxiv.org/abs/2508.02895.

\begin{figure}
\centering
\includegraphics[width=1.\columnwidth]{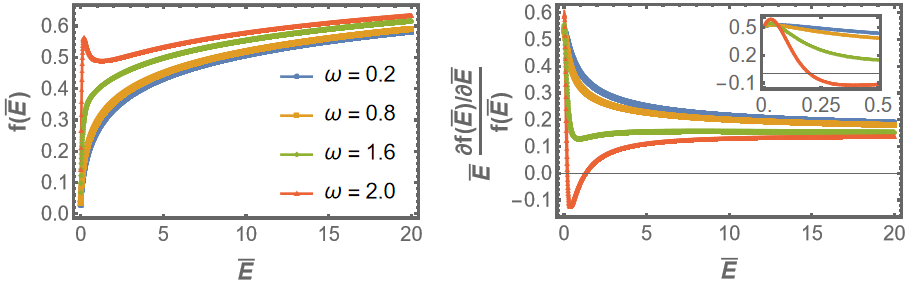}
\caption{\small{Numerical results for $\Delta=\Delta_+=5/2, l=0$ of 4-d holographic CFT. (a) Plots of the function $f(\bar{E})$ with different fixed values of $\omega$. (b) Plots of $\bar{E} f'(\bar{E})/f(\bar{E})=\mu f'(\mu)/f(\mu)$ with the different fixed values of $\omega$. $\bar{E}$-axes are in units of $1/\tilde{G}_N$.}}
\label{fEbvsEb_plots_ads5}
\end{figure}

Figure \ref{fEbvsEb_plots_ads5}(a) is plots of the function $f(\bar{E})$ for $\Delta=\Delta_+=5/2, l=0$, corresponding to $m^2=-15/4$ with different fixed values of $\omega$. It is clear that it is a monotonically increasing function for $\omega$ smaller and away from $\omega=\Delta$. It is true for the entire range of $r_h$, particularly for small black holes. For $\omega=2.0$, we can see that the monotonic behaviour is locally violated for certain range of $r_h$. This is due to the appearance of quasiparticles at frequency near $\Delta+l+2n, n\in \mathbb{N}_0$  for small black holes \cite{Jokela:2015sza}. Similar plots for $3-d$ CFT and non-zero $l$ in $4-d$ CFT are presented in Appendix \ref{app:num-res}.

The most important physical consequence of the monotonic nature of $f(\bar{E})$ is that a small black hole microstate will absorb more energy when perturbed over and above the entropic contribution, as one can see from (\ref{microDE}). $\beta\sim r_h$ for small $r_h$, so the entropic factor becomes less significant with decreasing $r_h$. Due to the very stiff nature of $f(\bar{E})$ at smaller $r_h$, smaller black hole microstates absorb more energy than the entropic contribution. Since we have taken particular values of $l$, it corresponds to be a global perturbation with the corresponding angular momentum. For example, $l=0$ corresponds to the global perturbation in \cite{Bhattacharyya:2009uu}. One can construct local perturbations by superimposing different $l$ modes.

Figure \ref{fEbvsEb_plots_ads5}(b) is plots of $\bar{E} f'(\bar{E})/f(\bar{E})=\mu f'(\mu)/f(\mu)$ for $\Delta=\Delta_+=5/2, l=0$ with the different fixed values of $\omega$. We can see that $\bar{E} f'(\bar{E})/f(\bar{E})$ is positive and finite. It tends towards a constant for both $\bar{E}\to 0$ and $\bar{E}\to \infty$. It is also mostly positive except for the quasiparticle dominant region. So, the 4-d holographic CFT satisfy R3 of ETH-monotonicity. The inset shows that $\bar{E} f'(\bar{E})/f(\bar{E}) \to \kappa=\kappa_0=0.5$ as $\bar{E}\to 0$. Moreover, the large $\bar{E}$ limit also tends to a constant which means that $\nu=1$ in (\ref{ETHmonoinf}).

\begin{figure}
\centering
\includegraphics[width=1.\columnwidth]{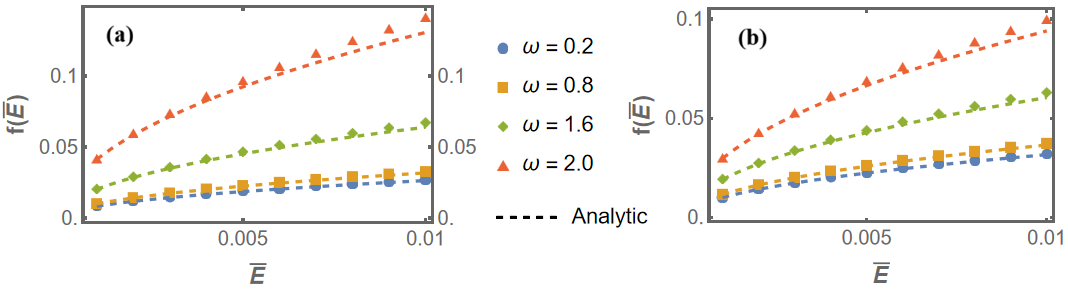}
\caption{\small{Comparison of $f(\bar{E})$ from numerics and the analytic expression (\ref{fl0exp}). Dashed lines are plots of the analytic expression for the different values of $\omega$. $\bar{E}$-axis is in unit of $1/\tilde{G}_N$. (a) is for $\Delta=\Delta_+=5/2, l=0$ of 4-d holographic CFT. (b) is for $\Delta=\Delta_+=3, l=0$ of 4-d holographic CFT.}}
\label{analytic-fEb}
\end{figure}

We can derive the analytic expression of $f(\bar{E})$ for $l=0$. We use the small $\mu$ expansion formula in \cite{Dodelson:2022yvn}. We refer to this work for the details. The retarded Green function takes the form
\begin{equation}
\frac{A\mu^{a}+B\mu^{-a}}{C\mu^{a}+D\mu^{-a}}, \qquad a=\frac{l+1}{2}
\end{equation}
$A$, $B$, $C$ and $D$ are various functions of $r_h\sim \sqrt{\mu}\sim 1/T$, $l$, $\Delta$, and $\omega$. The full expression is invariant under $a\to -a$. The leading order term $A/C$ has been studied in detail, but it is a real function in $\mu\to 0$ limit. So, the imaginary part, which is equal to the spectral function $A(\omega)$, appears only in the sub-leading terms starting with $\mu^{(l+1)}$. As mentioned in \cite{Dodelson:2022yvn}, these subleading terms which are non-perturbative in $l$ encode information of the presence of the horizon. So, it makes sense that we can extract curvature at the horizon from $f(\bar{E})$ which can be calculated only from these non-perturbative terms. For $l=0$, we look at the leading imaginary part of the Taylor expansion of
\begin{equation}
\mu\;\text{Im} \left[\;\lim_{r_h=\sqrt{\mu}\to 0}\left(\frac{B}{C}-\frac{A D}{C^2}\right)\right]
\end{equation}
This gives us the functional form in (\ref{fl0exp}). This expression of $f(\bar{E})$ matches our numerical results, as one can see in Figure \ref{analytic-fEb}. Figure \ref{analytic-fEb}(a) is plots for $\Delta=\Delta_+=5/2, l=0$ of 4-d holographic CFT. Similarly, Figure \ref{analytic-fEb}(b) is plots for $\Delta=\Delta_+=3, l=0$ of 4-d holographic CFT. We have used $\lim_{\Delta\to 3}\;\sin (\pi  \Delta )\, \Gamma (2-\Delta )=-\pi$.

In principle, one can obtain the analytic expressions of $f(\bar{E})$ for all $l>0$ in the limit  at $\mu\to 0$. One needs to take into account the instanton part of the Nekrasov-Shatashvili free energy. It is not necessary for $l=0$. The cancellation of the various divergent functions should also be performed with extreme care. We will not pursue this objective in the present work.

\begin{figure}
\centering
\includegraphics[width=1.\columnwidth]{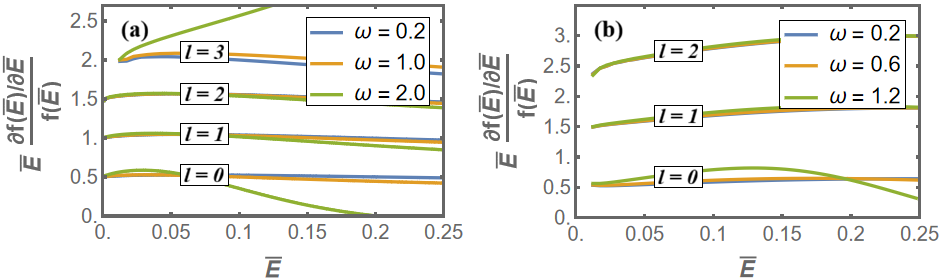}
\caption{\small{Plots of $\bar{E} f'(\bar{E})/f(\bar{E})=\mu f'(\mu)/f(\mu)$ as $\bar{E}\to 0$ considering different values of $l$ for (a) $4-d$ holographic CFT with $\Delta=\Delta_+=5/2$ and (b) $3-d$ holographic CFT with $\Delta=\Delta_+=7/4$. The limiting value $\kappa$ is given by $\kappa=\kappa_0+d_{\kappa}l$ with $\kappa_0=0.5$.}}
\label{kappa0}
\end{figure}

We analyse the limiting value $\kappa$ for different dimensions and different angular momentum modes $l$ in Figure \ref{kappa0}. Figure \ref{kappa0}(a) are plots of $\bar{E} f'(\bar{E})/f(\bar{E})$ for different values $l$ for $4-d$ holographic CFT with $\Delta=\Delta_+=5/2$. Figure \ref{kappa0}(b) is similar plots for $3-d$ holographic CFT with $\Delta=\Delta_+=7/4$. We can see that $\kappa$ varies as $\kappa=\kappa_0+d_{\kappa}l$ with $\kappa_0=0.5$. The plot for $l=2$ in $3-d$ case appears to deviate away from $2.5$ but we believe that it is part of an oscillation, as in the inset in Figure \ref{fEbvsEb_plots_ads5}(b). Numerical accuracy drops drastically with increasing value of $l$.

\begin{figure}
\centering
\includegraphics[width=1.\columnwidth]{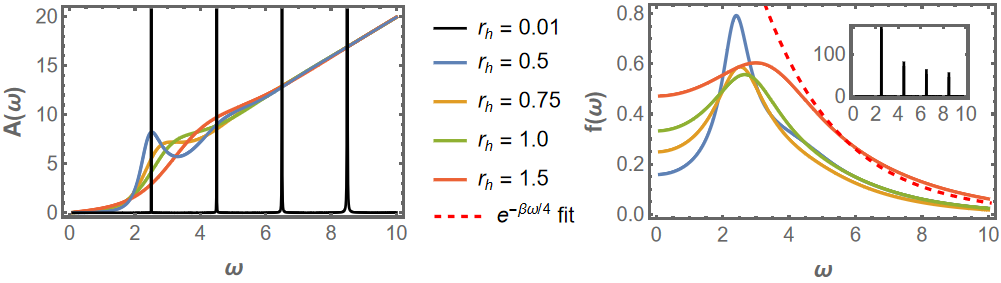}
\caption{\small{(a) Plots of the spectral function $A(\omega)$ for $\Delta=\Delta_+=5/2, l=0$ with different fixed values of $r_h$. (b) Plots of $f(\omega)$ for $\Delta=\Delta_+=5/2, l=0$ with the different fixed values of $r_h$.}}
\label{Aomega_fomega_plots_ads5}
\end{figure}

Figure \ref{Aomega_fomega_plots_ads5}(a) is plots of the spectral function $A(\omega)$ with different fixed values of $r_h$. We can see the appearance of quasiparticles for the very small black hole $r_h=0.01$. The spectral function has polynomial divergence as $\omega\to\infty$. It is the usual UV divergence of CFT. Figure \ref{Aomega_fomega_plots_ads5}(b) is plots of $f(\omega)$ with the different fixed values of $r_h$. We can see it mostly falls exponentially as $e^{-\beta\omega/4}$. It appears to be falling slightly slower due to the polynomial UV divergence.

If we look at the change in energy (\ref{thermalDE}) and (\ref{microDE}), UV divergence implies that the source $\lambda(t)$ cannot have sharp kinks or discontinuities. Similar observation has been made in \cite{Mandal:2015kxi} for free field CFTs where the sharp step function limit of quantum quenches cannot be taken.

We present further numerical results in Appendix \ref{app:num-res}. It includes results for non-zero $l$, results for $\mathcal{O}_{\Delta-}$ and results for the $3$-d CFT from $3+1$-D asymptotically AdS black hole.

\section{ETH-monotonicity in two-dimensional holographic CFT}
\label{2d-CFT}
The metric of the BTZ black hole \cite{Banados:1992wn} is
\begin{equation}
ds^2=-(-\mu+r^2)dt^2+\frac{dr^2}{(-\mu+r^2)}+r^2d\phi^2, \qquad \phi\in[0,2\pi)
\end{equation}
The CFT at the AdS boundary lives in $S^1\times R$ where $S^1$ is the spatial circle. The horizon radius, the Hawking temperature, the total energy and the Bekenstein-Hawking entropy are given by
\begin{equation}
r_h=\sqrt{\mu}, \qquad T=\frac{r_h}{2\pi},\qquad \bar{E}=M=\frac{\mu}{8G_N}=\frac{r_h^2}{\tilde{G}_N}, \qquad S=\frac{\pi r_h}{2 G_N}.
\end{equation}
The spectral function has been calculated in \cite{Son:2002sd}. One should take the momentum mode $k\in \mathbb{Z}$. The Green functions in $S^1\times R$ and $R\times R$ are different. Particularly in $S^1\times R$, one should identify $\phi =\phi +2\pi n, \; n\in \mathbb{Z}$ and sum over the image points \cite{Lifschytz:1993eb,Shiraishi:1993qnr,Steif:1993zv}. But in going to momentum basis, one can simply take the $n=0$ term and perform the $\phi$-integral from $-\infty$ to $+\infty$ which takes care of the image points. We again consider the zero momentum modes $k=0$. The spectral function is given by
\begin{gather}
A(\beta,\omega)=\frac{1}{\beta^{2\Delta-2}}\sinh\frac{\beta\omega}{2}\left|\Gamma\left(\frac{\Delta}{2}+\frac{i\beta\omega}{4\pi}\right)\right|^4
\end{gather}
Note that the $1/\beta^{2\Delta-2}$ factor comes from the normalization. The conformal weight $\Delta \geq 0$. This gives us
\begin{equation}
\eta=3/2, \qquad \kappa=\frac{\pi\omega}{4},
\end{equation}
in the ETH-monotonicity statement (\ref{ETHmono0}) for $\Delta$ equal to integers and half integers. $A(\beta,\omega)$ reduces to a product of elementary functions for integers values of $\Delta$.
\begin{equation}
A(\beta,\omega)=\begin{cases} \frac{1}{\beta^{2\Delta-2}}\,\frac{\pi\sinh\frac{\beta\omega}{2}}{\cosh^2\frac{\beta\omega}{4}}\left[\prod_{m=1}^{(\Delta-1)/2}\left(\left(m-\frac{1}{2}\right)^2+\frac{\beta^2\omega^2}{16\pi^2}\right)\right]^2 & \qquad \Delta\in \mathbb{N}\; \text{odd}\\
\frac{\omega^2}{16\beta^{2\Delta-4}}\,\frac{\sinh\frac{\beta\omega}{2}}{\sinh^2\frac{\beta\omega}{4}}\left[\prod_{m=1}^{\Delta/2-1}\left(m^2+\frac{\beta^2\omega^2}{16\pi^2}\right)\right]^2 & \qquad \Delta\in \mathbb{N}\; \text{even}\\
\end{cases}
\end{equation}
This gives the $f$-function in the canonical ensemble as
\begin{equation}
f(\beta,\omega)=\begin{cases}\frac{\pi}{\beta^{\Delta-1}}\,\text{sech}\frac{\beta\omega}{4}\prod_{m=1}^{(\Delta-1)/2}\left(\left(m-\frac{1}{2}\right)^2+\frac{\beta^2\omega^2}{16\pi^2}\right) & \qquad \Delta\in \mathbb{N}\; \text{odd}\\
\frac{\omega}{4\beta^{\Delta-2}}\,\text{cosech}\frac{\beta\omega}{4}\prod_{m=1}^{\Delta/2-1}\left(m^2+\frac{\beta^2\omega^2}{16\pi^2}\right) & \qquad \Delta\in \mathbb{N}\; \text{even}\\
\end{cases}
\end{equation}
The sech and cosech functions give the exponential suppression
\begin{equation}
e^{-\beta\omega/4}\sim e^{-\pi\omega/2r_h}\sim e^{-\pi\omega/(2\sqrt{\tilde{G}_N\bar{E}})}
\end{equation}
This is the exponential fall-off we have mentioned below (\ref{ETH}). In contrast, it reduces to one in higher-d CFTs due to negative specific heat. The exponential factor dominates the power law dependence on $\beta$ as $\bar{E}\to 0$. Hence, the changes in energy $\Delta E_{\beta}$ and $\Delta E_{mc}$ vanish exponentially in this limit. Note that $f(\bar{E})$ is an increasing function of $\bar{E}$. So, $2$-d CFTs still satisfy R2 of ETH-monotonicity.

We now check the contribution from quantum fluctuations $\Delta E_{mc,q}$. We calculate the quantity $\Delta E_{mc,q}/\Delta E_{\beta}$ for $2$-D CFT in the limit $\bar{E}\to 0$ or $r_h\to 0$.
\begin{equation}
\frac{\Delta E_{mc,q}}{\Delta E_{\beta}}=\frac{3}{8}\,\omega^2 S''(\bar{E})|_{\bar{E}=E_{mc}}=-3\pi G_N\,\frac{\omega^2}{\mu^{3/2}} = -\frac{3\pi \omega^2G_N}{r^{3}_h}
\label{Ediffratioq3D}
\end{equation}
We have used $S(\bar{E})=\pi\mu^{1/2}/(2 G_N)$ and $\bar{E}_{mc}=\mu/(8G_N)$. It is now a negative quantity because small BTZ black holes still have positive specific heat. On the other hand, $\overline{\Delta E}_{mc}$ is given by
\begin{equation}
\overline{\Delta E}_{mc} =\frac{\omega\kappa\tilde{G}_N}{\mu^{3/2}}= \frac{2\pi\omega^2 G_N}{r^3_h}
\end{equation}
So, quantum fluctuations numerically dominate ETH-monotonicity in $r_h\to 0$ limit, even though they are of the same order in $G_N$ and $r_h$.

The results of 2-d CFT are in agreement with the absence of curvature singularity in the BTZ black hole. Curvature invariants like the Kretschmann scalar is a fixed constant everywhere. We still find that $f(\bar{E})$ is monotonically increasing in the $2$-d holographic CFT. The quasiparticle peaks also do not appear in the limit $r_h\to 0$ because $r_h=0$ BTZ black hole is different from the $3$-D global AdS spacetime.

\section{Conclusions}
\label{cnd}
In this work, we study the the enveloping function $f(\bar{E},\omega)$ of the fluctuation term in the ETH statement (\ref{ETH}) for holographic conformal field theories. We use the $\bar{E}$-dependence of this function to identify and study black hole microstates. In particular, we are interested in the microstates of small black holes which have high curvature at the horizon. We also set down a set of criteria for the $\bar{E}$-dependence which we called ETH-monotonicity. It reinforces the second law of thermodynamics over and above the entropic contribution in microcanonical ensembles of finite systems. It is a manifestation of inequivalence of ensembles. We show that higher dimensional holographic CFTs possess ETH-monotonicity. So, smaller black hole microstates in higher dimensions have stronger contribution from ETH-monotonicity to the second law of thermodynamics. ETH-monotonicity also dominates other quantum fluctuations in small black hole microstates in higher dimensions. The black hole singularity is constructed of microstates for which ETH-monotonicity starts competing with the entropic factor.

We show that the relative extra gain in energy measures the curvature at the black hole horizon. This is expected in other black hole geometries with negative specific heat and satisfying ETH-monotonicity. Two-dimensional holographic CFT does not possess all features of ETH-monotonicity. The gain in energy vanishes exponentially with shrinking horizon radius. Other quantum fluctuations also negate and numerically dominate ETH-monotonicity in this case. These results are in agreement with the absence of curvature singularity in the BTZ black hole. So, we conclude that

\begin{center}
inequivalence of ensembles = ETH-monotonicity = curvature at the horizon
\end{center}
in higher dimensional holographic conformal field theories. We have studied all competing terms to come to this conclusion. We have quantified these relations by calculating the relative extra gain in energy (\ref{Ediffratioval}) from ETH-monotonicity. It also holds true for pure states within a microcanonical energy window. We expect that ETH-monotonicity will persist even in the ultimate quantum theory of gravity. Because it is a property of many-body quantum chaotic system which becomes more prominent with decreasing system size. In a large black hole, ETH-monotonicity got washed out by the large entropy, and correspondingly, the curvature at the horizon is small.

Following is the list of the various energy scales and the various competing quantities appearing in our calculations:
\begin{enumerate}
\item Energy is measured in units of $1/R$ where $R$ is the AdS radius. We have taken $R=1$. $\bar{E}$, $M$ and $S$ are order $O(1/G_N)$ quantities. The microcanonical energy window $\delta$ is order $O(1/\sqrt{G_N})$. $\mu$, $\omega$, $r_h$, $\beta$ and $T$ are order $O(1)$ quantities in appropriate units. The microcanonical energy fluctuation is relevant only for experimental measurements. We have full theoretical control at orders $O(1/G_N)$, $O(1)$ and $O(G^1_N)$. The most interesting physics is at order $O(G^1_N)$.
\item The gain in energy from the entropic contribution $\Delta E_{\beta}$ is order $O(1)$, as expected for perturbations with light operators. But it is parametrically much smaller than one because we take $\tilde{\lambda}(\omega)$ to be very small so that perturbative calculation works.
\item The extra gain in energy from ETH-monotonicity $\Delta_{mc,f}$ and the quantum fluctuations $\Delta E_{mc,q}$ are order $O(G^1_N)$. But we have shown that $\Delta E_{mc,q}$ is parametrically much smaller that $\Delta E_{mc,f}$ in small black holes in higher dimensions. So, ETH-monotonicity dominates the quantum fluctuations in this regime. Correction at the level of the microcanonical energy window $\delta$ are suppressed by $e^{-1/G_N}$, so it does not affect our analysis.
\item One may be wondering about the contribution of the quantum fluctuation in the change in energy starting from a canonical ensemble (\ref{thermalDE}). But we are not interested in this correction because we have taken the expression as it is. It has been called the entropic contribution. If one is still interested in this correction for large black holes with stable canonical ensemble, it can be easily calculated. It will be further suppressed by another factor of $G_N$.
\item When calculating $f(\bar{E})$ using (\ref{ETHspec}) in small black holes, one should ideally use microcanonical ensemble. It will introduce $\omega$ in the $\bar{E}$-dependence just like the ETH-monotonicity contribution in $\Delta E_{mc}$. But again, the overall correction from this minute difference will be order $O(G^2_N)$.
\end{enumerate}

One may be wondering about diverging temperature of small black holes in higher dimensions. Statistically it would imply that the probability density is equal for all microstates which does not make sense. After all, there is also another infinite temperature limit for large black holes. A speculative way out of this conundrum is provided in \cite{Asplund:2008xd}. It was proposed that the small black hole microstates are dual to eigenstates of a SU(M) subset of degrees of freedom of a SU(N) gauge theory. In the present work, we are simply performing a phenomenological analysis by measuring the temperature and the spectral function from the small black hole background.

We believe that our technique of using $f(\bar{E})$ to identify and study black hole microstates will have wide applications in black hole physics and, in general, in quantum gravity. It is order $O(G^1_N)$ effect but it can be tracked with precision. It is not unreasonable to expect that this idea would apply to black holes in asymptotically flat spacetime as well as small black holes in asymptotically de-Sitter spacetime. After all, small black holes have the same behaviour in any of these spacetimes. The only requirement is that the system is strongly chaotic so that ETH statement holds true. Finally, it is also very tempting to wonder if the Big Bang singularity were a very highly energetic large ``white hole", such as the one we defined in footnote \ref{whitehole}. 

\section*{Acknowledgement}
The author thanks Gautam Mandal and Diptarka Das for helpful discussions. The author is fully supported by the Department of Science and Technology (Government of India) under the INSPIRE Faculty fellowship scheme\\
(Ref. No. DST/INSPIRE/04/2020/002105).

\appendix

\section{Energy gain starting from a microcanonical ensemble}
\label{app:mc}
We will show that the extra gain in energy due to ETH-monotonicity persists in case of a microcanonical ensemble. For illustrative purpose, we use a toy density matrix constructed out of two eigenstates $|n_1\rangle$ and $|n_2\rangle$ with corresponding energy eigenvalues $E_1$ and $E_2$ respectively. Consider $E_1<E_2$. The density matrix is
\begin{equation}
\rho=\frac{1}{2}\left(|n_1\rangle\langle n_1|+|n_2\rangle\langle n_2|\right)
\end{equation}
The energy expectation value is $\bar{E}_{mc}=(E_1+E_2)/2$. $\delta=E_2-E_1$ is the energy window of the microcanonical ensemble. It is of order $O(1/\sqrt{G_N})$ but parametrically much smaller than $\sqrt{G_N}\bar{E}_{mc}$. Let us have a closer look at the Taylor expansion of the entropic terms in (\ref{microDEexact}) for the two eigenstates.  For simplicity let us consider for a single frequency $\omega$.
\begin{eqnarray}
&& S(E_{1}+\omega)-S(E_{1}+\omega/2)\nonumber\\
&=& S(\bar{E}_{mc}-\frac{\delta}{2}+\omega)-S(\bar{E}_{mc}-\frac{\delta}{2}+\omega/2)\nonumber\\
&=&\frac{\beta\omega}{2}-\frac{\omega\delta}{4}\,S''(\bar{E})|_{\bar{E}=\bar{E}_{mc}}+\frac{3\omega^2}{8}\,S''(\bar{E})|_{\bar{E}=\bar{E}_{mc}}+\frac{\omega\delta^2}{16}\,S'''(\bar{E})|_{\bar{E}=\bar{E}_{mc}}+O(G^{2}_N)\nonumber\\
&& S(E_{2}+\omega)-S(E_{2}+\omega/2)\nonumber\\
&=& S(\bar{E}_{mc}+\frac{\delta}{2}+\omega)-S(\bar{E}_{mc}+\frac{\delta}{2}+\omega/2)\nonumber\\
&=&\frac{\beta\omega}{2}+\frac{\omega\delta}{4}\,S''(\bar{E})|_{\bar{E}=\bar{E}_{mc}}+\frac{3\omega^2}{8}\,S''(\bar{E})|_{\bar{E}=\bar{E}_{mc}}+\frac{\omega\delta^2}{16}\,S'''(\bar{E})|_{\bar{E}=\bar{E}_{mc}}+O(G^{3/2}_N)\nonumber\
\end{eqnarray}
The quantum fluctuations term $\omega^2 S''$ has been treated separately in section \ref{qf}. Taylor expansion of the exponentials of the remaining terms will kill all the odd $\delta$ power terms. It appears that terms like $(\omega\delta S'')^2$ and $\omega\delta^2S'''$ are order $O(G^1_N)$ quantities which could be a matter of concern. But all these terms with $\delta$ dependence will be suppressed by $e^{-1/G_N}$. This is because we take $M=O(e^{1/G_N})$ microstates in a generic microcanonical ensemble. So, the probability of each microstate is $1/M$. Any factor of $\delta/2$ should be then replaced by the difference of the energy eigenvalue from the microcanonical energy expectation value. For example, the final contribution from the first non-trivial term from the Taylor expansion of $e^{\omega\delta^2 S'''/16}$ is
\begin{equation}
\frac{\omega\delta^2S'''}{4M}\sum_{m=1}^M\frac{1}{m^2}=\frac{\omega\delta^2S'''}{M}\frac{\pi^2}{6} \sim O(e^{-1/G_N}).
\end{equation}
We have taken large $M$ limit because we are working with many-body excited states. Now let us consider the ETH-monotonicity form factor, again for the toy density matrix for easier illustration.
\begin{eqnarray}
&&\omega |\tilde{\lambda}(\omega)|^2\,e^{\beta\omega/2} \left[f(E_{1}+\omega/2,\omega)^2+f(E_{2}+\omega/2,\omega)^2\right]\nonumber\\
&=&\omega |\tilde{\lambda}(\omega)|^2\,e^{\beta\omega/2} \left[f(\bar{E}_{mc}-\delta+\omega/2,\omega)^2+f(\bar{E}_{mc}+\delta+\omega/2,\omega)^2\right]\nonumber\
\end{eqnarray}
The $\delta$-dependent term starts from $(f'^2+ff'')\delta^2$. In a generic microcanonical ensemble, it is again suppressed by $e^{-1/G_N}$.

\section{Further numerical results}
\label{app:num-res}
\begin{figure}
\centering
\includegraphics[width=1.\columnwidth]{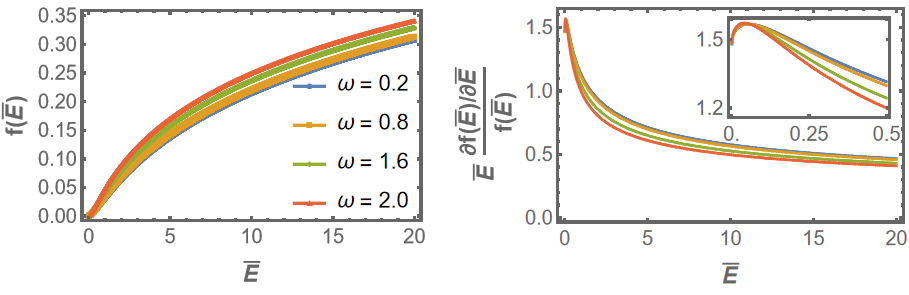}
\caption{\small{Numerical results for $\Delta=\Delta_+=5/2, l=2$ of 4-d holographic CFT. (a) Plots of the function $f(\bar{E})$ with different fixed values of $\omega$. (b) Plots of $\bar{E} f'(\bar{E})/f(\bar{E})$ with the different fixed values of $\omega$.}}
\label{fEbvsEb_plots_ads5_L2}
\end{figure}

Figure \ref{fEbvsEb_plots_ads5_L2}(a) is plots of the function $f(\bar{E})$ for $\Delta=\Delta_+=5/2$, corresponding to $m^2=-15/4$, but with $l=2$. In this case, the quasiparticles will appear at $\omega=\Delta+l+2n, n\in \mathbb{N}_0$. Figure \ref{fEbvsEb_plots_ads5_L2}(b) is plots of $\bar{E} f'(\bar{E})/f(\bar{E})$ for $\Delta=\Delta_+=5/2$ with the same fixed values of $\omega$. We again find that $\bar{E} f'(\bar{E})/f(\bar{E})$ is positive and finite. It is also a constant in the limits $\bar{E}\to 0$ and $\bar{E}\to \infty$.

\begin{figure}
\centering
\includegraphics[width=1.\columnwidth]{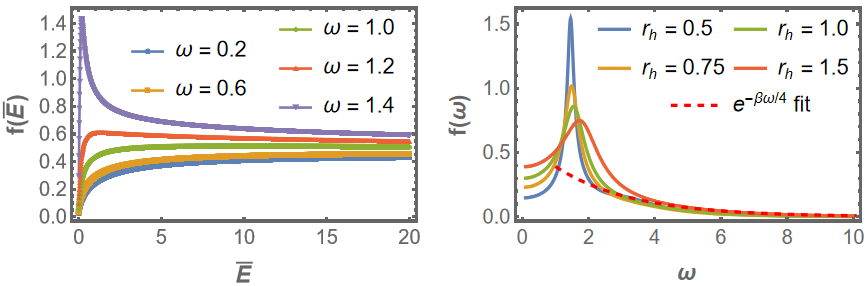}
\caption{\small{Numerical results for $\Delta=\Delta_-=3/2, l=0$ of 4-d holographic CFT. (a) Plots of the function $f(\bar{E})$ with different fixed values of $\omega$. (b) Plots of $f(\omega)$ with different fixed values of $r_h$.}}
\label{AbyB_Delta3half}
\end{figure}

Figure \ref{AbyB_Delta3half}(a) is plots of the function $f(\bar{E})$ for $\Delta=\Delta_-=3/2, l=0$, corresponding to $m^2=-15/4$. We have used (\ref{AbyBformula}) for this calculation. We again find that $f(\bar{E})$ is a monotonically increasing function of $\bar{E}$, in general. In this case, the quasiparticles will appear at $\omega=\Delta_-+l+2n, n\in \mathbb{N}_0$. Figure \ref{AbyB_Delta3half}(b) is plots of the $f$-function with varying $\omega$ for $\Delta_-=3/2, l=0$ with different fixed values of $r_h$. We can see the quasiparticle peak prominently for smaller value of $r_h$. We can also see the exponential fall-off of $e^{-\beta\omega/4}$.

\begin{figure}
\centering
\includegraphics[width=1.\columnwidth]{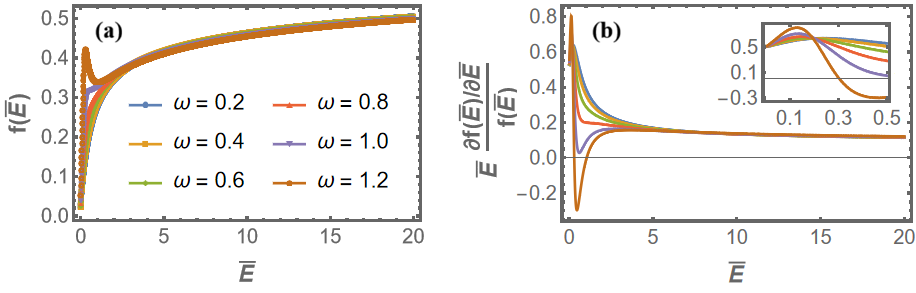}
\caption{\small{Numerical results for $\Delta=\Delta_+=7/4, l=0$ of 3-d holographic CFT. (a) Plots of the function $f(\bar{E})$ with different fixed values of $\omega$. (b) Plots of $\bar{E} f'(\bar{E})/f(\bar{E})$ with the different fixed values of $\omega$.}}
\label{ads4_Delta7fourth}
\end{figure}

Figure \ref{ads4_Delta7fourth} are numerical results for the 3-d holographic CFT from $(3+1)$-D asymptotically AdS black hole. We again find that the scalar operator satisfies ETH-monotonicity.

\begin{figure}
\centering
\includegraphics[width=1.\columnwidth]{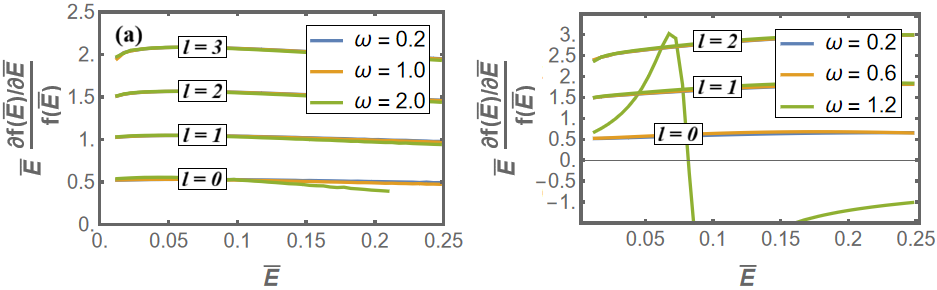}
\caption{\small{Plots of $\bar{E} f'(\bar{E})/f(\bar{E})=\mu f'(\mu)/f(\mu)$ as $\bar{E}\to 0$ considering different values of $l$ for (a) $4-d$ holographic CFT with $\Delta=\Delta_-=3/2$ and (b) $3-d$ holographic CFT with $\Delta=\Delta_-=5/4$. The limiting value $\kappa$ is again given by $\kappa=\kappa_0+d_{\kappa}l$ with $\kappa_0=0.5$.}}
\label{kappa0_3half}
\end{figure}

Figure \ref{kappa0_3half}(a) is plots of $\bar{E} f'(\bar{E})/f(\bar{E})$ for $\Delta=\Delta_-=3/2$ with different values of $l$ for 4-d holographic CFT. Figure \ref{kappa0_3half}(b) is plots of $\bar{E} f'(\bar{E})/f(\bar{E})$ for $\Delta=\Delta_-=5/4$ with different values of $l$ for 3-d holographic CFT. The wild variation for $l=0,\; \omega=1.2$ is because of the quasiparticle peak at $\omega=5/4=1.25$.

\bibliography{refs}
\bibliographystyle{JHEP}

\end{document}